# Zero Poisson's Ratio and Suppressed Mechanical Anisotropy in BP/SnSe Van der Waals Heterostructure: A First-principles Study

Qi Ren[a], Xingyao Wang[a], Yingzhuo Lun[a], Xueyun Wang[a], and Jiawang Hong[a,*]

[a] School of Aerospace Engineering, Beijing Institute of Technology, Beijing, 100081, China

[*] Corresponding author E-mail: hongjw@bit.edu.cn (J. Hong)

**Abstract**

Black phosphorene and its analogs have attracted intensive attention due to their unique puckered structures, anisotropic characteristics, and negative Poisson's ratio. The van der Waals heterostructures assembly by stacking different materials may show novel physical properties which the parent materials don't possess. In this work, the first-principles calculations were performed to study the mechanical properties of the BP/SnSe van der Waals heterostructure. Interestingly, a near-zero Poisson's ratio $v_{zx}$ was found in BP/SnSe heterostructure. In addition, compared to the parent materials BP and SnSe with strong in-plane anisotropic mechanical properties, the BP/SnSe heterostructure shows strongly suppressed anisotropy. Our findings suggest that the vdW heterostructure could show quite different mechanical properties from the parent materials and provide new opportunities for the mechanical applications of the heterostructures.

## 1. Introduction

Two-dimensional (2D) materials such as graphene[1], hexagonal boron nitride (*h*-BN)[2], and transition metal dichalcogenides (TMDs)[3] have received widespread attention owing to their extraordinary structural, mechanical, and electronic properties[4,5]. Among them, black phosphorene (BP) has been widely studied due to its unique in-plane anisotropy[6,7], negative Poisson's ratio[8], infrared bandgap energy[9,10], and high carrier mobility[11,12], *etc*. These properties render BP a promising 2D material for polarization sensors[13], plasmonic devices[14], field-effect transistor[15], *etc*. Recently, BP-like analogues such as GeS, GeSe, SnS, and SnSe have been extensively investigated due to similar puckered structure and anisotropic characteristics[16–18]. Moreover, the advantages of high thermoelectric figure-of-merit[19,20], earth abundance[21], environmental compatibility[22], and less toxicity[23] make them a fertile playground for exploring potential photovoltaic and thermoelectric devices[24,25].

Stacking 2D van der Waals (vdW) materials, namely constructing vdW heterostructures, has opened new avenue for fundamental scientific studies and device designs. These vdW heterostructures provide unique properties such as atomically thin transistors[26], gate-tunability[27], and superconductivity[28,29], imbuing them with more novel functionalities than their parent materials. Recently, BP has been regarded as a new candidate material for 2D vdW heterostructures, and the preparation[30], lattice structures[31], mechanical properties[32], *etc*. of BP-based heterostructures have been gradually studied. For example, the BP/SnSe vdW heterostructure has been investigated in terms of its potential applications in photovoltaic devices[33], but little is known about its mechanical properties. Therefore, it is necessary to investigate the mechanical properties of BP/SnSe vdW heterostructure for its application and the durability of the heterostructure-based devices.

Herein, the mechanical properties, including the ideal strength, critical strain, Young's modulus, and Poisson's ratio of BP/SnSe vdW heterostructure, were investigated using the first-principles methods. The parent materials, i.e., BP and SnSe monolayers, was found to possess high anisotropic of the in-plane elastic modulus[6,16] and the novel negative Poisson's ratio[8,34]. Surprisingly, our results show that the in-plane anisotropy of mechanical properties in BP/SnSe heterostructure is much lower than that of the parent materials. In addition, the negative Poisson's ratio disappears and the BP/SnSe heterostructure shows a nearly zero Poisson's ratio results from unique thickness variation of the BP/SnSe heterostructure. These results suggest that mechanical properties of vdW heterostructure can be much different from the constructing materials, and therefore the mechanical design of the heterostructure-based devices should be carefully performed.

## 2. Methods

The density functional theory (DFT) calculation was carried out using the projector augmented wave (PAW)[35] scheme with the Perdew–Burke–Ernzerhof (PBE) functional of generalized gradient approximation (GGA)[36] method as implemented in the Vienna *ab initio* simulation package (VASP)[37,38]. A plane wave cutoff of 450 eV was set in our calculations. K-point samplings of 8 × 10 × 1 (slab model) and 8 × 10 × 3 (bulk model) were used. DFT-D2 level[39] was used in our calculations to taking into consideration of the van der Waal forces. Atomic relaxation was performed until the force on each atom is smaller than 0.005 eV Å$^{-1}$, and the total energy change was less than $10^{-4}$ eV.

## 3. Results and discussion

### 3.1 Geometric Structures

The crystal structures of BP, SnSe and two kinds of BP/SnSe vdW heterostructure are shown in Figure 1. The lattice constants of BP and SnSe were calculated, which are in consistent with previous reported experimental and theoretical values[6,40–42], as listed in Table 1. It can be seen that the lattice constant *a* (zigzag) slightly decreases and *b* (armchair) increases in BP from bulk to monolayer, while the trend of SnSe is opposite. Note that the lattice constant *a* of the BP/SnSe vdW heterostructure is larger than that of bilayer BP and smaller than that of bilayer SnSe, and the lattice constant *b* is larger than both of them. Two different stackings of BP and SnSe, AA-stacking (top layer directly stacks on the bottom layer) and AB-stacking (top layer shifts half unit cell), were considered. It shows that the AB-stacking is energetically more favorable than the AA-stacking structure. Therefore, AB-stacking was chosen in our subsequent study.

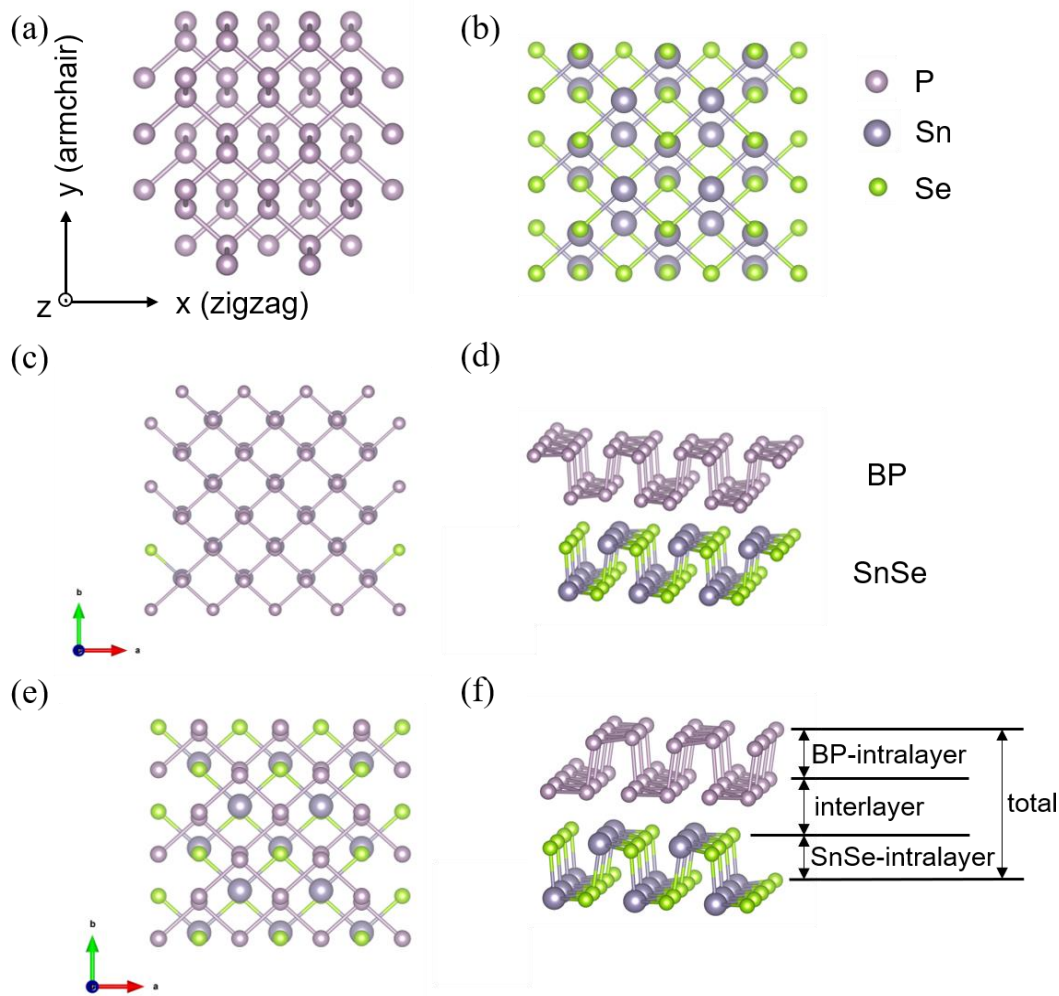

**Figure 1.** Atomic structures of BP (a) and SnSe (b), AA-stacking of BP and SnSe heterostructure with top view (c) and side view (d), AB-stacking with top view (e) and side view (f).

**Table 1.** The lattice constants of monolayer (1L), bilayer (2L) and bulk of the BP, SnSe, and BP/SnSe vdW heterostructure

| Material | | Lattice constant (Å) | |
|---|---|---|---|
| | | a (zigzag) | b (armchair) |
| BP | 1L | 3.305 | 4.564 |
| | 2L | 3.312 | 4.511 |
| | bulk | 3.321 | 4.429 |
| | | (3.313[40], | (4.374[40], |
| | | 3.308[6]) | 4.536[6]) |
| SnSe | 1L | 4.244 | 4.313 |
| | 2L | 4.224 | 4.322 |
| | bulk | 4.190 | 4.374 |
| | | (4.153[41], | (4.445[41], |
| | | 4.22[42]) | 4.36[42]) |
| BP/SnSe | 2L | 3.632 | 4.575 |

### 3.2 Mechanical Anisotropy

The stress-strain curves along zigzag and armchair directions (in-plane) of BP, SnSe, and BP/SnSe vdW heterostructure were calculated, as shown in Figure 2. The Young's modulus was obtained by fitting the stress-strain curve in the linear region of small strain, as summarized in Table 2. It shows that the Young's modulus along zigzag and armchair direction for monolayer, bilayer and bulk are nearly the same for BP and SnSe. The Young's modulus of BP is about 4 times larger than that of SnSe along zigzag direction and 3 times larger along armchair direction, due to the stronger P-P bonds than Sn-Se bonds. Along both zigzag and armchair directions, the Young's modulus of BP/SnSe vdW heterostructure is larger than that of the bilayer SnSe and smaller than that of the bilayer BP. In addition, both BP and SnSe show strong anisotropy of Young's modulus, i.e., the anisotropic ratio is 4.0 for bilayer BP and 3.0 for bilayer SnSe, which origins from their puckered structure[8]. Surprisingly, the anisotropic ratio of Young's modulus of BP/SnSe vdW heterostructure structure is much weaker than that of the parent materials, i.e., it is only 1.7. This is because the Young's modulus along armchair direction has significant enhancement (2.2 times) compared to SnSe, while it only has 1.2 times enhancement along zigzag direction. These results indicate that the mechanical properties of vdW heterostructure can be much different from the parent materials.

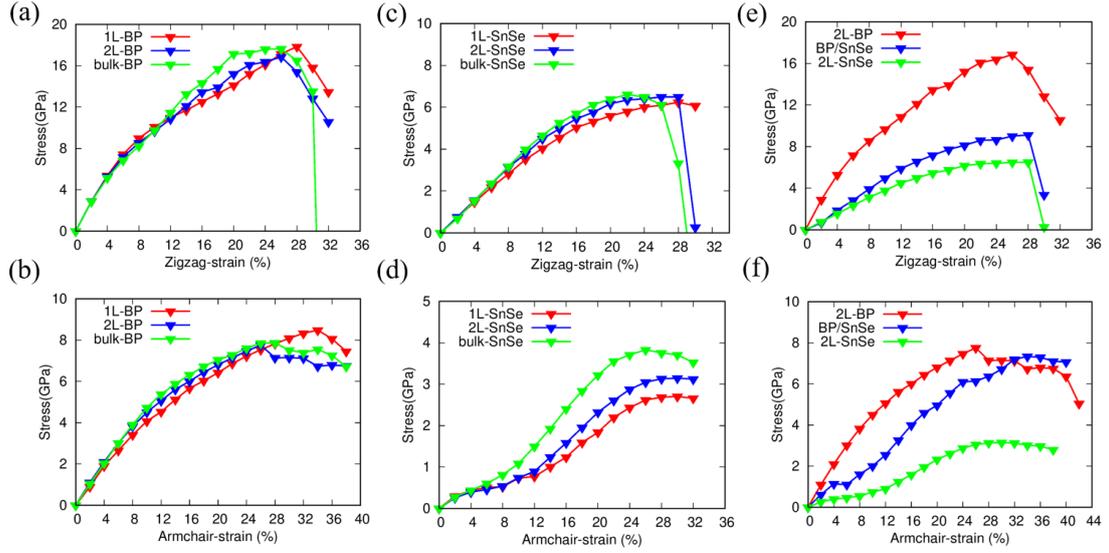

**Figure 2.** Comparisons of the stress-strain curves of monolayer, bilayer, and bulk of BP (a-b), SnSe (c-d), and BP/SnSe heterostructure (e-f) with different tensile directions. Top panels: zigzag direction, bottom panels: armchair direction.

**Table 2.** Young's modulus of BP, SnSe and BP/SnSe heterostructure, the anisotropic ratio is the ratio of properties along zigzag over that of armchair direction

|  |  | Young's modulus (GPa) | | Anisotropic |
| --- | --- | --- | --- | --- |
|  |  | Zigzag (x) | Armchair (y) | ratio |
|  | 1L | 163.21 | 42.47 | 3.8 |
| BP | 2L | 160.33 | 40.08 | 4.0 |
|  | bulk | 156.92 | 39.48 | 4.0 |
|  | 1L | 36.76 | 10.82 | 3.4 |
| SnSe | 2L | 38.26 | 12.93 | 3.0 |
|  | bulk | 39.44 | 13.89 | 2.8 |
| BP-SnSe | 2L | 47.02 | 28.47 | 1.7 |

The ideal strength is the maximum achievable stress of a defect-free crystal and the critical strain is the strain at which the ideal strength reaches[43]. Both were obtained from the peak of the stress-strain curves in Fig.2, and were summarized in Table 3. Similar to the Young's modulus, the ideal strength and critical strain along zigzag and armchair direction for monolayer, bilayer and bulk are nearly the same for BP and SnSe. The ideal strength of BP is about 3 (2) times larger than that of SnSe along zigzag (armchair) direction, while the critical strains are very close for both materials. Interestingly, both BP and SnSe show strong anisotropy of ideal strength, i.e., the anisotropic ratio is 2.2 for BP and 2 for SnSe, while the critical strain shows very weak anisotropy.

**Table 3.** Ideal strength and critical strain of BP, SnSe and BP/SnSe vdW heterostructure, the anisotropic ratio is the ratio of properties along zigzag over that of armchair direction

|  |  | Ideal strength (GPa) | | | Critical strain (%) | | |
|---|---|---|---|---|---|---|---|
|  |  | zigzag | armchair | Anisotropic ratio | zigzag | armchair | Anisotropic ratio |
| BP | 1L | 17.84 | 8.48 | 2.10 | 28 | 34 | 0.82 |
|  | 2L | 16.85 | 7.75 | 2.17 | 26 | 26 | 1.00 |
|  | bulk | 17.62 | 7.86 | 2.24 | 26 | 28 | 0.93 |
| SnSe | 1L | 6.23 | 2.70 | 2.31 | 28 | 30 | 0.93 |
|        | 2L   | 6.50 | 3.15 | 2.06 | 28 | 30 | 0.93 |
|        | bulk | 6.61 | 3.83 | 1.73 | 22 | 26 | 0.85 |
| BP/SnSe | 2L  | 9.12 | 7.33 | 1.24 | 28 | 34 | 0.82 |

The stress-strain curves of BP/SnSe vdW heterostructure along zigzag and armchair direction are both between the curves of bilayer BP and SnSe, as shown in Figure 2(e) and (f). However, the stress-strain curve of BP/SnSe heterostructure along zigzag direction is closer to that of the bilayer SnSe, and it is similar to bilayer BP along armchair direction. Interestingly, the anisotropic ratio of ideal strength of BP/SnSe is much weaker than that of the parent materials, i.e, it is only 1.2 times for heterostructure while it is about 2 for BP and SnSe. Unlike ideal strength, the critical strain of BP/SnSe is similar to those of bilayer of BP and SnSe.

### 3.3 Zero Poisson's ratio

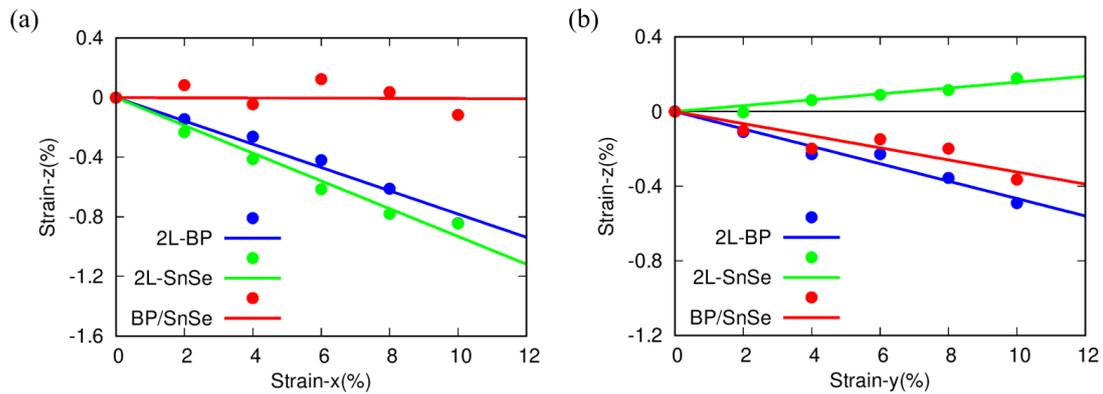

**Figure 3.** The out-of-plane strain vs in-plane strain along zigzag (x) direction (a) and armchair (y) direction (b) for bilayer BP, SnSe, BP/SnSe vdW heterostructure.

**Table 4**. Poisson's ratio of bilayer BP, SnSe, and BP/SnSe heterostructure

| Materials | $v_{zx}$ | $v_{zy}$ |
|-----------|----------|----------|
| BP        | 0.078    | 0.047    |
| SnSe      | 0.093    | -0.016   |
| BP/SnSe   | 0.001    | 0.032    |

Figure 3 shows the out-of-plane strain (*z* direction) under uniaxial deformation along zigzag (*x*) and armchair (*y*) directions in the range from 0 % to 10 %. Poisson's ratio is defined by $v_{ij} = - \varepsilon_i / \varepsilon_j$, where $\varepsilon_j$ is the uniaxial strain applied in the *j* axis and $\varepsilon_i$ is the resulting strain along the *i* axis. The results for the Poisson's ratios were summarized in Table 3. It shows that the bilayer of BP and SnSe possess normal Poisson's ratio except for $v_{zy}$ of SnSe, which is negative value. Surprisingly, after combing monolayer of SnSe and BP to form the vdW heterostructure, its Poisson's ratio is strikingly different from the bilayer BP and SnSe, i.e., the out-of-plane strain of BP/SnSe is almost unchanged with in-plane strain along zigzag direction, indicating the nearly zero Possions' ratio of $v_{zx}$.

To understand the origins of the differences between the Poisson's ratios of bilayer BP, SnSe and BP/SnSe, we examine the change of their intralayer, interlayer and the total thickness (each thickness can be found in Fig.1f) under uniaxial strain $\varepsilon_x$ and $\varepsilon_y$. It can be seen that thickness change of intralayer and interlayer always has opposite sign under tensile strain along

$x$ or $y$ direction, both for bilayer BP and SnSe. Interestingly, the $v_{zy}$ shows slight negative Poisson's ratio due to the increase of the intralayer thickness is larger than the decrease of the interlayer, as shown in Fig.6d. As for BP/SnSe heterostructure under tensile strain along the $x$-direction (Fig.6e), the intralayer thickness of SnSe layer increases while the intralayer of BP keeps nearly constant. Interestingly, the interlayer thickness between SnSe and BP monolayer decreases with nearly the same amount of SnSe intralayer thickness change, resulting the total thickness of heterostructure nearly unchanged with tensile along $x$ direction, i.e., zero Poisson's ration $v_{zy}$. Note that the intralayer thickness of SnSe and the interlayer thickness exhibit a sudden change at around 6% strain in heterostructure along $y$ direction (Fig.6f), due to the change of SnSe structure as revealed in previous research[34].

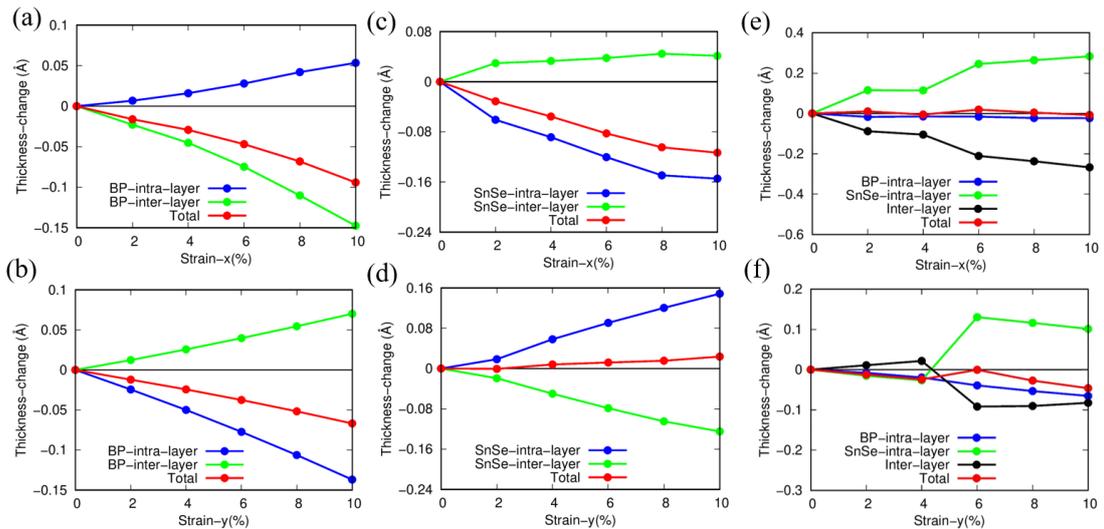

**Figure 4.** The change of thickness of bilayer BP (a-b), SnSe (c-d), and BP/SnSe heterostructure (e-f) with different tensile directions.

## 4. Conclusion

In summary, the mechanical properties of BP, SnSe, and BP/SnSe vdW heterostructure were investigated using the first-principles methods. Our results indicate that the vdW heterostructure can have quite different mechanical properties from the parent materials. A near-zero Poisson's ratio $v_{zx}$ was found in BP/SnSe heterostructure origins from its unique thickness variation. Besides, the in-plane anisotropy of mechanical properties in BP/SnSe heterostructure is much lower than that of the parent materials BP and SnSe. The ideal strength of the BP/SnSe vdW heterostructure in both zigzag and armchair direction are higher than that of SnSe but lower than that of BP. Therefore, by stacking different 2D materials to form the vdW heterostructure, it may offer novel mechanical properties and greatly extend the applications of the parent materials.

**Acknowledgement**

This work is supported by the National Natural Science Foundation of China (Grant Nos. 11572040, 11604011), the National Key Research and Development Program of China (Grant No. 2019YFA0307900), Beijing Natural Science Foundation (Grant No. Z190011), and the Technological Innovation Project of Beijing Institute of technology. Theoretical calculations were performed using resources of the National Supercomputer Centre in Guangzhou.